\documentclass[12pt]{article}

\usepackage{amsmath, amssymb, slashed}
\usepackage[dvips]{graphicx}
\usepackage{epsf}
\usepackage{color}

\def\b{\beta}

\begin{document}

%%%%%%%%%%%%%%%%%%%%%%
%%%%% Title page %%%%%
%%%%%%%%%%%%%%%%%%%%%%
\begin{titlepage}
\begin{center}

\hfill IPMU12-0186\\
\hfill ICRR-report-629-2012-18\\
%\hfill \today

\vspace{1.5cm}

{\large \bf Heavy Squarks and Light Sleptons in Gauge Mediation} \\
\vspace{2.5mm}
{\large \sl From the viewpoint of 125 GeV Higgs Boson and Muon $g-2$}

\vskip 1.2cm

{Masahiro Ibe}$^{(a, b)}$,
{Shigeki Matsumoto}$^{(b)}$, \\
{Tsutomu T. Yanagida}$^{(b)}$,
and
{Norimi Yokozaki}$^{(b)}$

\vskip 0.4cm

{\it
$^{(a)}${\it ICRR, University of Tokyo, Kashiwa, 277-8583, Japan} \\
$^{(b)}${Kavli IPMU, University of Tokyo, Kashiwa, 277-8583, Japan}
}

\vskip 1.5cm

\abstract{
In the framework of gauge mediation models, we investigate scenarios with heavy squarks and light sleptons, motivated by the recent discovery of the Higgs boson and the deviation  of the muon anomalous magnetic moment ($g-2$) from the SM prediction. We show that only models with a messenger multiplet in the adjoint representation of $SU(5)$ GUT gauge group are the unique possibility that sleptons are light enough to explain the muon $g-2$ in the minimal setup. We also show that, if there is an additional source of the Higgs soft masses, the muon $g-2$ can be explained with messenger multiples in the fundamental representation of $SU(5)$ with the help of the light higgsino. Some phenomenological aspects of these models are also discussed.
}

\end{center}
\end{titlepage}

\setcounter{footnote}{0}

%%%%%%%%%%%%%%%%%%%%%%%%
%%%%% Introduction %%%%%
%%%%%%%%%%%%%%%%%%%%%%%%
\section{Introduction}
\label{sec: introduction}

The latest results from the ATLAS and CMS collaborations show a $5\sigma$ signal for a Higgs-like boson mass at around 125 GeV~\cite{ATLAS_Higgs, CMS_Higgs}. The results have significant impacts on the model building of the supersymmetric (SUSY) standard model (SSM). For example, a lightest Higgs boson mass around 125 GeV requires large stop masses of $\mathcal{O}(10-100)$ TeV~\cite{OYY1, OYY2} (see, e.g.,~\cite{PureGM} for recent discussion) or large stop $A$-term~\cite{OYY2} in the minimal SSM (MSSM). In other cases, the 125 GeV Higgs boson mass requires some extensions of the MSSM, such as models with massive extra matter fields coupling to the Higgs doublets~\cite{extramatter}, models with extra gauge interactions~\cite{extragauge}, or models with a singlet fields coupling to the Higgs doublet, e.g., the next-to-minimal SSM (NMSSM) (see \cite{NMSSM1, NMSSM2} for recent reviews).%~\footnote{More drastic extensions with strong interactions have also been proposed~\cite{Drastic}.}

On another front, the seeming discrepancy of the muon anomalous magnetic moment ($a_\mu$) between theoretical predictions and experimental results has been a strong motivation to expect that superparticles are discovered in the near future at the LHC experiments. In the SSM, the discrepancy can be resolved when the superparticles, especially sleptons as well as the higgsino, wino and bino are in the mass range of $\mathcal{O}(100)$ GeV. Such expectations are, however, getting disappointed (especially in the MSSM) due to the unexpectedly heavy Higgs boson mass which tends to require the superparticle masses in the TeV range or above.

In response to this situation, it is imperative to investigate whether or not the 125 GeV Higgs boson mass and the discrepancy of the anomalous magnetic moment can be explained simultaneously in the models of the SSM. It has been shown that both can be achieved in the MSSM with gauge mediation where a large $A$-term is generated due to the mixing between Higgs doublets and messenger fields~\cite{EIY_gm2}. It has also been shown that both can also be achieved simultaneously in the extensions of the MSSM with extra matter fields~\cite{EXT_gm2},\footnote{The higgsino mass of $\mathcal{O}(1)$ TeV as well as the SUSY masses of the extra matters can be related to the PQ-symmetry breaking scale~\cite{PQ_EXT}.} or extra gauge interactions~\cite{EXT_gauge_gm2}.

In this paper, we discuss an alternative possibility to explain the Higgs boson mass of 125 GeV and $a_\mu$ in models with heavy squarks and light sleptons based on the models with gauge mediation. In this scenario, the Higgs boson mass is explained by heavy stop masses at $\mathcal{O}(10)$ TeV, while the discrepancy of $a_\mu$ is explained by light sleptons. As we will see later, this apparently effortless possibility is highly constrained. As a result, we find that the model with messenger fields in the adjoint {\bf 24} representation of the $SU(5)$ gauge group of the grand unified theory (GUT) is the unique possibility to explain both Higgs boson mass and $a_\mu$ by the heavy squark and the light slepton spectrum. We also show that the models with messengers in fundamental ${\bf 5}+\bar{\bf 5}$ representations can also explain the Higgs boson mass and $a_\mu$, once we admit extra contributions to the Higgs soft-mass squared.

This paper is organized as follows. In section \ref{sec: heavysquark}, we discuss generic conditions to explain the Higgs boson mass and $a_\mu$ on models with heavy squarks and light sleptons. In section \ref{sec: adjoint}, we discuss the model with messenger fields in the adjoint representation. In section\,\ref{sec: addH}, we discuss models with additional Higgs soft-mass squared. The final section is devoted to summary of our discussions.

%%%%%%%%%%%%%%%%%%%%
%%%%% Muon g-2 %%%%%
%%%%%%%%%%%%%%%%%%%%
\section{Muon ${\bf g-2}$ vs. heavy squarks}
\label{sec: heavysquark}

The muon anomalous magnetic moment ($a_\mu = g - 2$) has been measured very precisely, and it is an important probe of new physics beyond the standard model (SM). The current experimental value of the anomalous magnetic moment is~\cite{Bennett:2006fi}:
\begin{eqnarray}
a^{\rm exp}_\mu = 11659208.9(6.3) \times 10^{-10}.
\end{eqnarray}
The most recent calculation of the SM prediction, on the other hand, is~\cite{Hagiwara:2011af};
\begin{eqnarray}
a_\mu^{\rm SM}= 11659182.8(4.9) \times 10^{-10},
\end{eqnarray}
which includes the updated data from "$e^+ e^- \to$ hadrons" processes and the latest evaluation of the hadronic light-by-light scattering contributions. As a result, the experimental result and the SM prediction deviate with each other by about $3.3\sigma$,
\begin{eqnarray}
\delta a_\mu = a_\mu^{\rm exp} - a_\mu^{\rm SM}
= (26.1\pm 8.0) \times 10^{-10}.
\end{eqnarray}

In the MSSM, $\delta a_{\rm \mu}$ can be resolved by supersymmetric contributions when the sleptons (smuons) as well as the higgsino and/or the wino and the bino are in the range of $\mathcal{O}(100)$ GeV. In our scenario, the stop masses are expected to be $\mathcal{O}(10)$ TeV to explain the 125 GeV Higgs boson mass.\footnote{It is worth noting that $A$-terms are suppressed in models with gauge mediation. As discussed in references \cite{EIY1, EIY_gm2}, sizable $A$-terms can be achieved by mixing messenger fields and Higgs doublets even in models with gauge mediation, though we do not peruse in this paper.} To achieve such heavy squark masses, we first assume that messenger fields with color charges generate
\begin{eqnarray}
m_{\rm squark}
\sim \frac{g_3^2}{16\pi^2} \frac{F_c}{M_c}
= \mathcal{O}(10) \, {\rm TeV},
\label{eq:squark}
\end{eqnarray}
where $g_3$ denotes the $SU(3)_C$ gauge coupling constant, while $M_c$ and $F_c$ are the mass and mass splitting of the colored messenger, respectively. The slepton masses are, on the other hand, expected to be of $\mathcal{O}(100)$ GeV for explaining the observed anomalous magnetic moment,
\begin{eqnarray}
m_{\rm slepton}
\sim \frac{g_{1, 2}^2}{16\pi^2} \frac{F_w}{M_w}
= \mathcal{O}(100) \, {\rm GeV},
\label{eq:slepton}
\end{eqnarray}
where $g_1$ and $g_2$ denote the gauge couplings of $U(1)_Y$ and $SU(2)_L$ (the SM gauge group), while $M_w$ and $F_w$ are a mass and a mass splitting of the non-colored messenger, respectively. In the following discussions, we assume that the mass scales of the colored and non-colored messengers are close with each other, i.e. $M_c \simeq M_w \simeq M$, for simplicity.
%so that the unification of gauge coupling constants remains intact.%
The mass splittings are, on the other hand, different from each other, and assumed to be hierarchical, i.e., $F_w \ll F_c$. Such different mass splittings are, for example, realized when couplings between the messenger fields and a supersymmetry breaking field depend on the fields whose vacuum expectation values break the GUT spontaneously. Notice that the different mass splitting $ F_w \ll F_c$ we choose in the following discussion does not disturb the unification of the SM gauge coupling constants.

The above required separation between the squark and slepton masses as in Eqs.\,(\ref{eq:squark}) and (\ref{eq:slepton}) is, however, not easily realized.
% while keeping perturbative unification of the gauge coupling constants at the GUT scale. 
For example, when the messenger is "{\it a pair of fundamental and anti-fundamental representations}", the colored messenger has non-vanishing $U(1)_Y$ charge, and hence, the right-handed sleptons obtain at least
\begin{eqnarray}
m_{\rm slepton} \sim \frac{3 }{5 \sqrt{2}}\frac{g_1^2}{g_3^2}  \times m_{\rm squark}.
%m_{\rm slepton} \sim \frac{6g_1^2}{5g_3^2} \times m_{\rm squark},∑
\end{eqnarray}
The right-handed sleptons cannot be light as $\mathcal{O}(100)$ GeV for $m_{\rm squark} = \mathcal{O}(10)$ TeV. Another complication comes from the size of the $\mu$-term. Due to the stop masses of $\mathcal{O}(10)$ TeV, the Higgs soft-mass squared ($m_{H_{u,d}}^2$) receive large radiative corrections,
\begin{eqnarray}
{\mit\Delta} m_{H_u}^2
\simeq \frac{6y_t^2}{16\pi^2} m_{\rm stop}^2 \log \frac{m_{\rm stop}^2}{M^2}
= -\left[ (3{\rm -}4) \, {\rm TeV} \right]^2.
\label{eq:radiative}
\end{eqnarray}
Here, $y_t$ denotes the coupling constant of the top Yukawa interaction. In order to realize the $Z$-boson mass ($m_Z$), the $\mu$-term is therefore required to satisfy
\begin{eqnarray}
\mu^2
\simeq \frac{m_{H_d}^2 - m_{H_u}^2 \tan^2\b}{\tan^2\b-1} - \frac{m_Z^2}{2}
\simeq -{\mit\Delta} m_{H_u}^2
= \left[ (3{\rm -}4) \, {\rm TeV} \right]^2,
\label{eq:mu}
\end{eqnarray}
where the Higgs mixing parameter $\tan\b$ is assumed to be large enough, $\tan\b = \mathcal{O}(10)$, to resolve the discrepancy of $a_\mu$ (see following discussions). We have also used $m_{H_u}^2 \simeq {\mit\Delta} m_{H_u}^2$, which is justified since the gauge mediated contributions to the Higgs soft-mass squared are the same as those of the sleptons, and hence, much smaller than ${\mit\Delta} m_{H_u}^2$. 

When the higgsino is heavy, a dominant contribution to the anomalous magnetic moment comes from the one-loop diagram in which the bino, left-, and right-handed sleptons are circulating, which is evaluated to be
\begin{eqnarray}
\delta a_\mu
\simeq\frac{3}{5} \frac{g_1^2}{8\pi^2}
\frac{m_\mu^2 \mu \tan\beta}{M_1^3} F_b,
\label{eq:g2}
\end{eqnarray}
where $m_\mu$ is the muon mass, $M_1$ the bino mass, and $F_b = \mathcal{O}(1)$ for $M_1 \simeq m_{\tilde \mu_R} \simeq m_{\tilde \mu_L}$~\cite{gm2_MSSM}. In order to resolve the discrepancy of $a_\mu$, the bino, left-, and right-handed sleptons are therefore required to be light as less than about 200-500 GeV, depending on the size of the left-right mixing, $\mu \tan\beta$. Most of colored messengers, however, have non-vanishing U(1)$_Y$ charges as mentioned above, which prevents the sleptons and bino from being light enough. One exception is the messenger in color octet representation which is embedded in the adjoint ${\bf 24}$ representation of the $SU(5)$ GUT gauge group. In this case, we can freely separate the squark and  slepton masses, and hence, the desired spectrum can be obtained. As we will see in the next section, the discrepancy of $a_\mu$ can be actually reduced in models with the adjoint messenger, while it is difficult in models with messengers in other representations. Note that too large $\mu\tan\beta$ is not allowed, since otherwise the electroweak symmetry breaking minimum becomes unstable. Therefore, $\delta a_\mu$ is  bounded from above by the stability constraint (see discussions in the next section).

A possible way to avoid the above conclusion is to introduce additional contributions to the Higgs soft square masses other than the gauge mediated contributions. With such extra contributions, the radiative contributions to the Higgs soft square masses from the heavy stops in Eq.~(\ref{eq:radiative}) can be fine-tuned to allow much smaller $\mu$-term. In such cases, we have additional supersymmetric contributions to $a_\mu$ from the one-loop diagrams in which the light higgsinos are circulating. As we will show in section \ref{sec: addH}, the discrepancy of $a_\mu$ can be resolved with not very light sleptons; the light higgsino and wino enhance the SUSY contributions to the muon $g-2$. In this way, models with messengers in other than the adjoint representations can also explain both the 125 GeV Higgs boson mass and the discrepancy of $a_\mu$.

%%%%%%%%%%%%%%%%%%%%%%%%%%%%%
%%%%% Adjoint messenger %%%%%
%%%%%%%%%%%%%%%%%%%%%%%%%%%%%
\section{Messenger in adjoint representation}
\label{sec: adjoint}

As we have discussed in previous section, the messenger in adjoint ${\bf 24}$ representation gives the unique possibility to have light sleptons in separation with squark masses. The adjoint ${\bf 24}$ messenger fields consist of $\Sigma_8$, $\Sigma_3$, $X$ and $\bar{X}$, which are, respectively, transformed as $({\bf 8},{\bf 1},0)$, $({\bf 1},{\bf 3},0)$, $({\bf 3}, {\bf 2},-5/6)$ and $({\bf \bar{3}}, {\bf 2}, 5/6)$ under the Standard model gauge groups $SU(3)_C$  $\times$ $SU(2)_L$ $\times$ $U(1)_Y$. The superpotential is given by
\begin{eqnarray}
W = (M_8 + k_8 F \theta^2) \, {\rm Tr} \Sigma_8^2
+ (M_3 + k_3 F \theta^2) \, {\rm Tr} \Sigma_3^2
+ (M_X + k_X F \theta^2) \, X \overline{X}.
\end{eqnarray}
We parameterize the SUSY breaking masses of the messengers by $k_8$, $k_3$ and $k_X$. Since only relative size is important, we take $k_8=1$ in the following analysis. Formulas for the soft mass parameters are given in Appendix A.

%%%%%%%%%%%%%%%%%%%%%%%%%%%%%%%%%%%%%%%%%%%%%%%%%%%%%%%%%%%%
\begin{figure}[t]
\begin{center}
\includegraphics[scale=0.52]{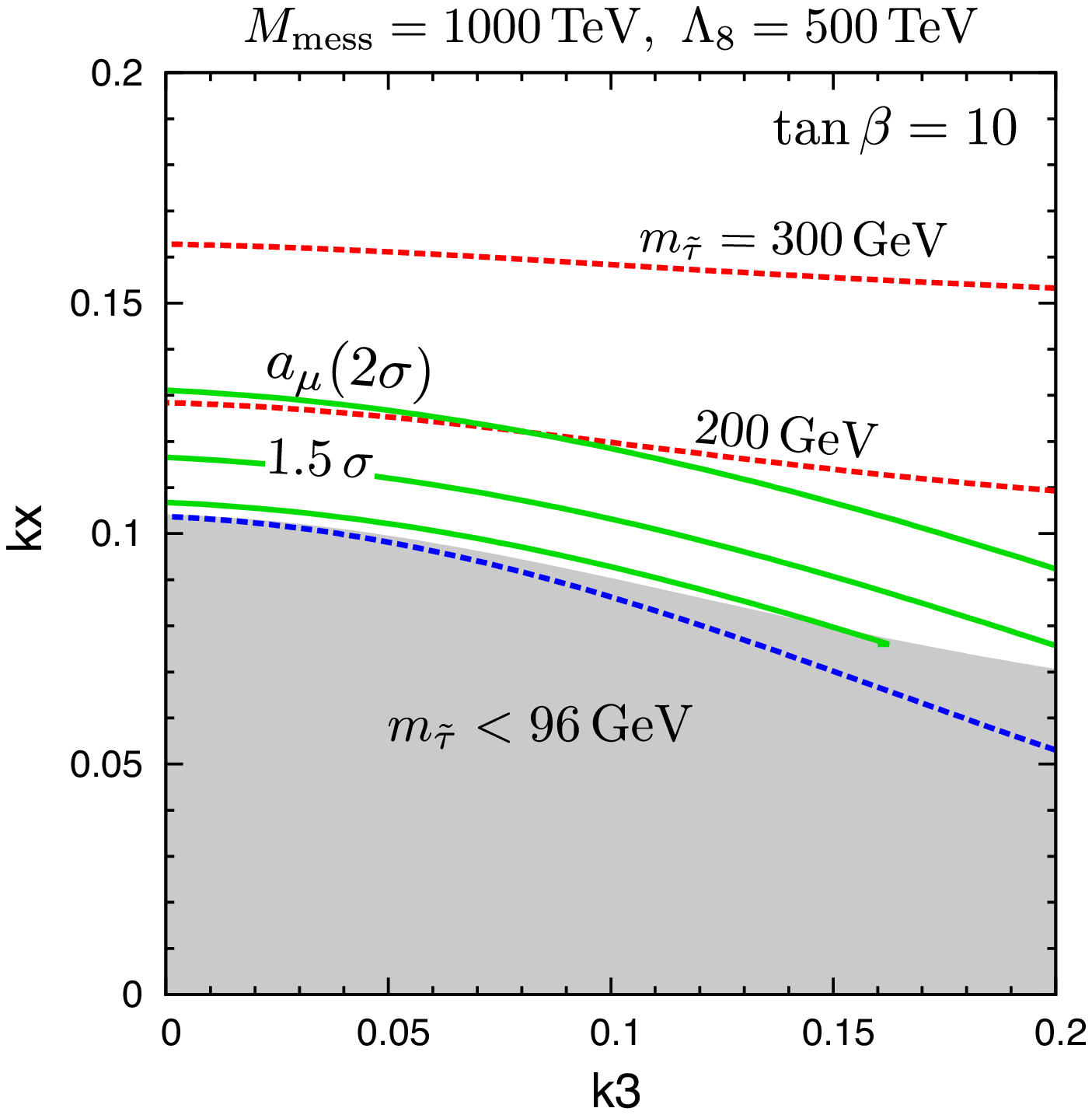}
\includegraphics[scale=0.52]{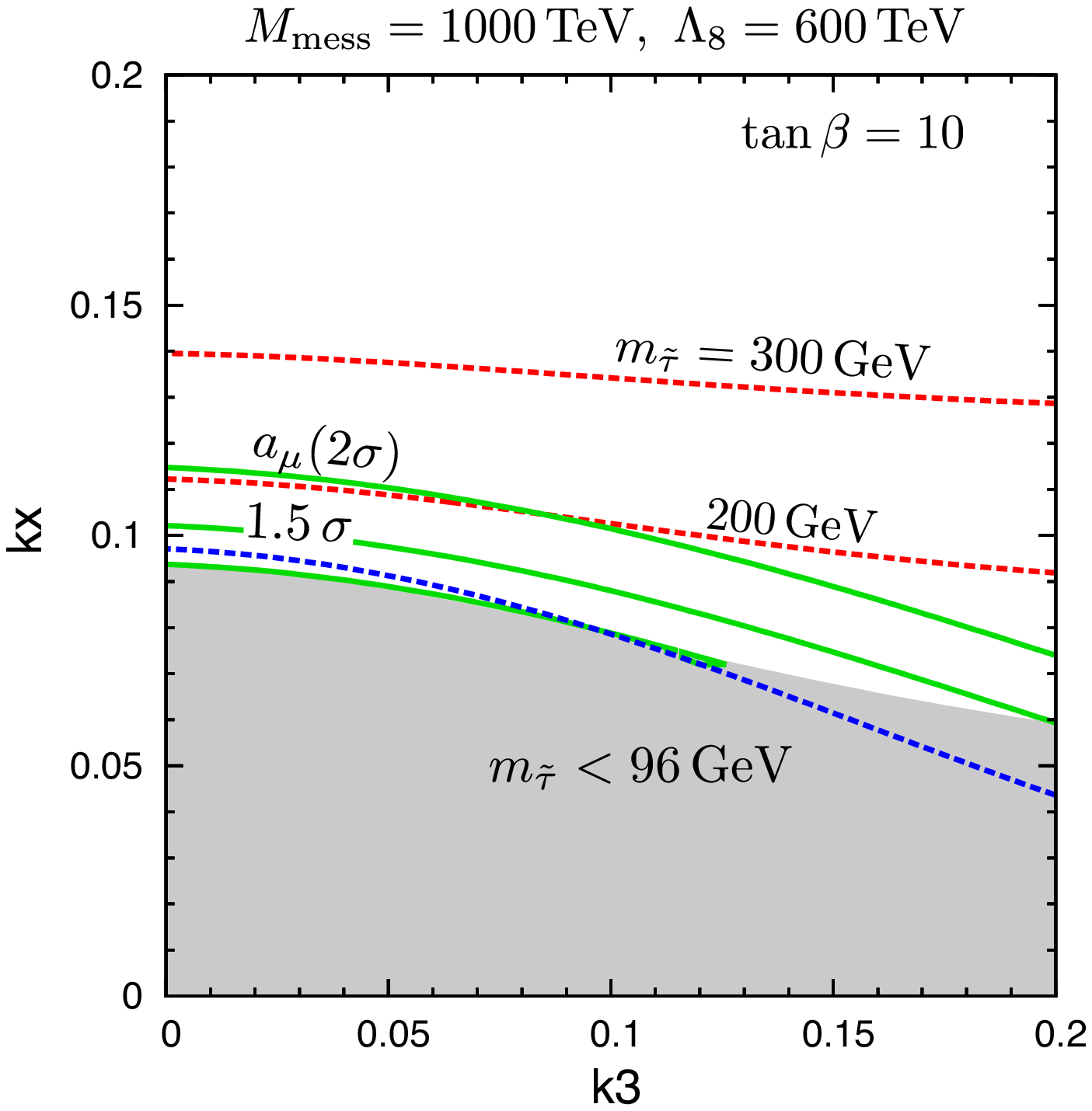}
\caption{\small \sl Contours of $a_\mu$ consistent with the experimental result at 2, 1.5, and 1$\sigma$ C.L. (from top to bottom) on the $k_3$-$k_X$ plane for $\Lambda_8 = $ 500 (left) and 600 TeV (right). The region below the dashed blue line is excluded because the electroweak symmetry minimum is not stable enough.}
\label{fig:24_1}
\end{center}
\end{figure}
%%%%%%%%%%%%%%%%%%%%%%%%%%%%%%%%%%%%%%%%%%%%%%%%%%%%%%%%%%%%

%%%%%%%%%%%%%%%%%%%%%%%%%%%%%%%%%%%%%%%%%%%%%
\begin{table}[t!]
  \begin{center}
    \begin{tabular}{  c | c  }
    $M_{\rm mess}$ & 1000 TeV \\
    $\Lambda_8$ & 500 TeV \\
%    $\tan\beta$ & 10 \\
    $k_3$ & 0.15 \\
    $k_X$ & 0.08 \\
    \hline
\hline    
    $M_{\rm bino}$ & 286 GeV \\
    $M_{\rm wino}$ & 691 GeV \\
    $\mu$ & 3.7 TeV \\
    \hline
    $m_{\rm gluino}$ & 9.9 TeV \\
    $m_{\tilde{t}}$ & 8.2 TeV \\
    $m_{\tilde{q}}$ & 8.8 TeV \\
    $m_{\tilde{e}_L} (m_{\tilde{\mu}_L})$ & 460 GeV\\
    $m_{\tilde{e}_R} (m_{\tilde{\mu}_R})$ & 170 GeV \\
    $m_{\tilde{\tau}_1}$ & 97 GeV \\
       $m_{\chi_1^0}$ & 277 GeV \\
     $m_{\chi_1^{\pm}}$ & 730 GeV \\
    $ \delta a_{\mu}$ & 1.80 $\times$ 10$^{-9}$
    \end{tabular}
    \hspace{20pt}
        \begin{tabular}{  c | c  }
    $M_{\rm mess}$ & 1000 TeV \\
    $\Lambda_8$ & 600 TeV \\
%    $\tan\beta$ & 10 \\
    $k_3$ & 0.16 \\
    $k_X$ & 0.07 \\
    \hline
\hline    
    $M_{\rm bino}$ & 301 GeV \\
    $M_{\rm wino}$ & 812 GeV \\
    $\mu$ & 4.3 TeV \\
    \hline
    $m_{\rm gluino}$ & 11.9 TeV \\
    $m_{\tilde{t}}$ & 9.7 TeV \\
    $m_{\tilde{q}}$ & 10.4 TeV \\
    $m_{\tilde{e}_L} (m_{\tilde{\mu}_L})$ & 549 GeV\\
    $m_{\tilde{e}_R} (m_{\tilde{\mu}_R})$ & 178 GeV \\
    $m_{\tilde{\tau}_1}$ & 113 GeV \\
      $m_{\chi_1^0}$ & 292 GeV \\
     $m_{\chi_1^{\pm}}$ & 857 GeV \\
    $ \delta a_{\mu}$ & 1.47 $\times$ 10$^{-9}$
    \end{tabular}
    \caption{\small \sl The mass spectrum and $\delta a_{\rm \mu}$ for two different points. The bino mass,  wino mass and higgsino mass parameters, defined at the scale, $m_{\tilde{t}}$, are also shown. As in Fig.~{\ref{fig:24_1}}, $\tan\beta=10$.
     }
  \label{table:24}
  \end{center}
\end{table}
%%%%%%%%%%%%%%%%%%%%%%%%%%%%%%%%%%%%%%%%%%%%%

In figure~\ref{fig:24_1}, several contours of $a_\mu$ are shown on the $k_3$-$k_X$ plane for the adjoint messengers with $\tan\beta = 10$. For simplicity, we take a common SUSY mass for the messengers, i.e., $M_8 = M_3 = M_X = M_{\rm mess}=10^6 $ GeV. The mass spectrum and renormalization group evolution are evaluated using {\tt SuSpect}~\cite{suspect} with appropriate modifications, while $\delta a_{\mu}$ (the SUSY contribution to $a_\mu$) is calculated by {\tt FeynHiggs}~\cite{FeynHiggs}. In the left (right) panel, $\Lambda_8 = F/M_8$ is taken to be $\Lambda_8 = 500\, (600)$ TeV, corresponding to the stop mass of $m_{\tilde{t}} \equiv (m_{\tilde{t}_1} m_{\tilde{t}_2})^{1/2} \simeq 8\,(10)$ TeV. In the region below the dashed blue line is excluded due to the unstable electroweak symmetry breaking minimum~\cite{Rattazzi:1996fb, Hisano:2010re}. This is because the large left-right mixing of the stau generates a deep charge breaking minimum, making the electroweak symmetry breaking minimum meta-stable. We have evaluated the stability bound using the fitting formula presented in reference~\cite{Hisano:2010re}. In both panels, we find a parameter region consistent with the experimental result of the muon $g-2$ at around 1\,$\sigma$ level. Remarkably, the result can be explained almost 1$\sigma$ level when the stop mass is about 8 TeV. It is worth noting that the constraint from the stability bound is severer for larger $\tan\beta$. No region consistent with the muon $g-2$ result can be found for larger $\tan\beta$, such as $\tan\beta=15$. In table~\ref{table:24}, we show the part of the  mass spectrum and $\delta a_\mu$  for two different points. Both left- and right-handed sleptons (smuons) as well as the bino are lighter than about $500$ GeV, which are required to explain the muon $g-2$, while the $\mu$ parameter is large as about 4 TeV. The colored SUSY particles are as heavy as $8-10$ ($10-12$) TeV in the left (right) column. Notice that the right-handed sleptons are even lighter than 200 GeV in the region of the parameter space consistent with the muon $g-2$ result within $1-1.5$ $\sigma$ level.

Due to the large $\mu$-parameter and moderately large $\tan\beta$, the stau always tends to be very light in this scenario. In fact, in the region consistent with the experimental result of the muon $g-2$, the stau is the next-to-lightest SUSY particle with its mass $m_{\tilde{\tau}} \lesssim 200$ GeV. When the R-parity is conserved, the stau is expected to decay into a gravitino by emitting a tau lepton with a long decay length (lifetime times the speed of light) and it can be regarded as a stable particle at collider experiments. Such a long-lived particle is, unfortunately, severely constrained to be $m_{\tilde{\tau}} >$ 270 GeV~\cite{ATLAS_stau1}. We therefore need some mechanisms to make the stau decay promptly.

One of the simplest ways to have the  prompt decay is use of the lepton number (R-parity) violating interactions, which are given by the superpotential,\footnote{One might worry about the existence of baryon number interactions too, but it is known that there are models which does not have such interactions with being consistent with GUT~\cite{LFV_models}.}
\begin{eqnarray}
W_{L \neq 0} = \lambda_{ijk} L_i L_j \bar{E}_k + \lambda'_{ijk} L_i Q_j \bar{D}_k,
\end{eqnarray}
where we take the basis of superfields so that the bilinear term $L H_u$ vanishes in the superpotential. The typical decay length of the stau is then estimated as
\begin{eqnarray}
c \tau_{\tilde{\tau}} \sim
\mathcal{O}(0.1 \, {\rm cm})
\left[ \frac{(\lambda, \lambda')}{10^{-7}} \right]^{-2}
\left( \frac{m_{\tilde{\tau}}}{100 \, {\rm GeV}} \right)^{-1}.
\label{eq:LFV_decay}
\end{eqnarray}
It is better to assume that the stau decays dominantly into a tau lepton and a neutrino, otherwise the stau mass tends to be constrained again by two jet resonance searches (through $L_i Q_j \bar{D}_k$ interactions) or searches using multi-lepton (electron/muon) channels at the LHC experiment. When the stau decays into a tau lepton and a neutrino, collider limits turn out to be very weak. Because all colored particles (gluino/squarks) are as heavy as $\mathcal{O}(10)$ TeV in this scenario, no limits on the stau mass has been obtained yet. Only the bound on the stau mass is from the LEP experiment, which gives the limit of $m_{\tilde{\tau}} >$ 95.9 GeV~\cite{RPVLEP}.

The couplings $\lambda$ and $\lambda'$ which are much larger than $\mathcal{O}(10^{-7})$ are not favored, since otherwise the baryon asymmetry produced in the early epoch of the universe would be washed out~\cite{washout}. The decay length of the stau therefore turns out definitely to be around $\mathcal{O}(mm)$, though it is hard to measure the corresponding impact parameter at the LHC experiment because of the small production cross section of the stau pair. Furthermore, if the lepton number is conserved in each flavor, one of $\lambda_{ijk}$ ($\lambda'_{ijk}$) can be larger than $\mathcal{O}(10^{-7})$ without causing the washout of the baryon asymmetry~\cite{LFV_LNV_ENDO}, which leads to much shorter decay length of the stau than Eq.(\ref{eq:LFV_decay}). In fact, lepton flavor violating terms do not arise in gauge mediation models, since the soft SUSY breaking masses of sleptons can be generated at a rather low-scale. It is also worth noting that, if the decay length is of the order of $mm$, the corresponding decay length can be observed at a future linear collider such as the ILC and CLIC~\cite{Matsumoto:2011fk}.

Another interesting possibility to have the prompt decay of the stau is the introduction of other particles which can be decay products of the stau. For example, when the axino (supersymmetric partner of the axion) exists within appropriate mass range, the stau can decays into a tau lepton and the axino. Then, the decay length of the stau can be small enough if the PQ breaking scale is as low as $\sim 10^9$ GeV.

%%%%%%%%%%%%%%%%%%%%%%%%%%%%
%%%%% Other messengers %%%%%
%%%%%%%%%%%%%%%%%%%%%%%%%%%%
\section{Additional Higgs soft-mass squared}
\label{sec: addH}

%%%%%%%%%%%%%%%%%%%%%%%%%%%%%%%%%%%%%%%%%%%%%%%%%%%%%%%%%%%%
\begin{figure}[t]
\begin{center}
\includegraphics[scale=0.51]{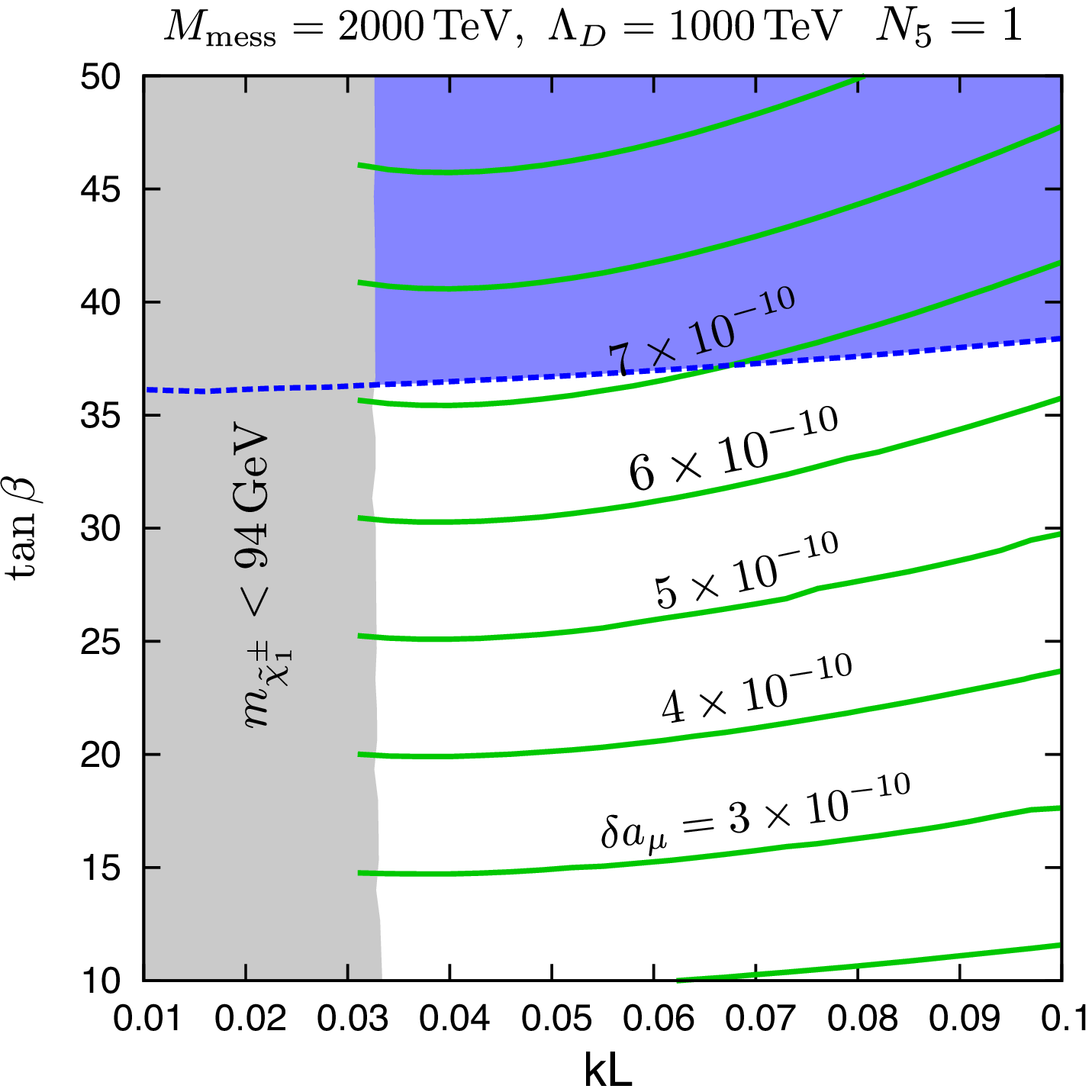}
\includegraphics[scale=0.51]{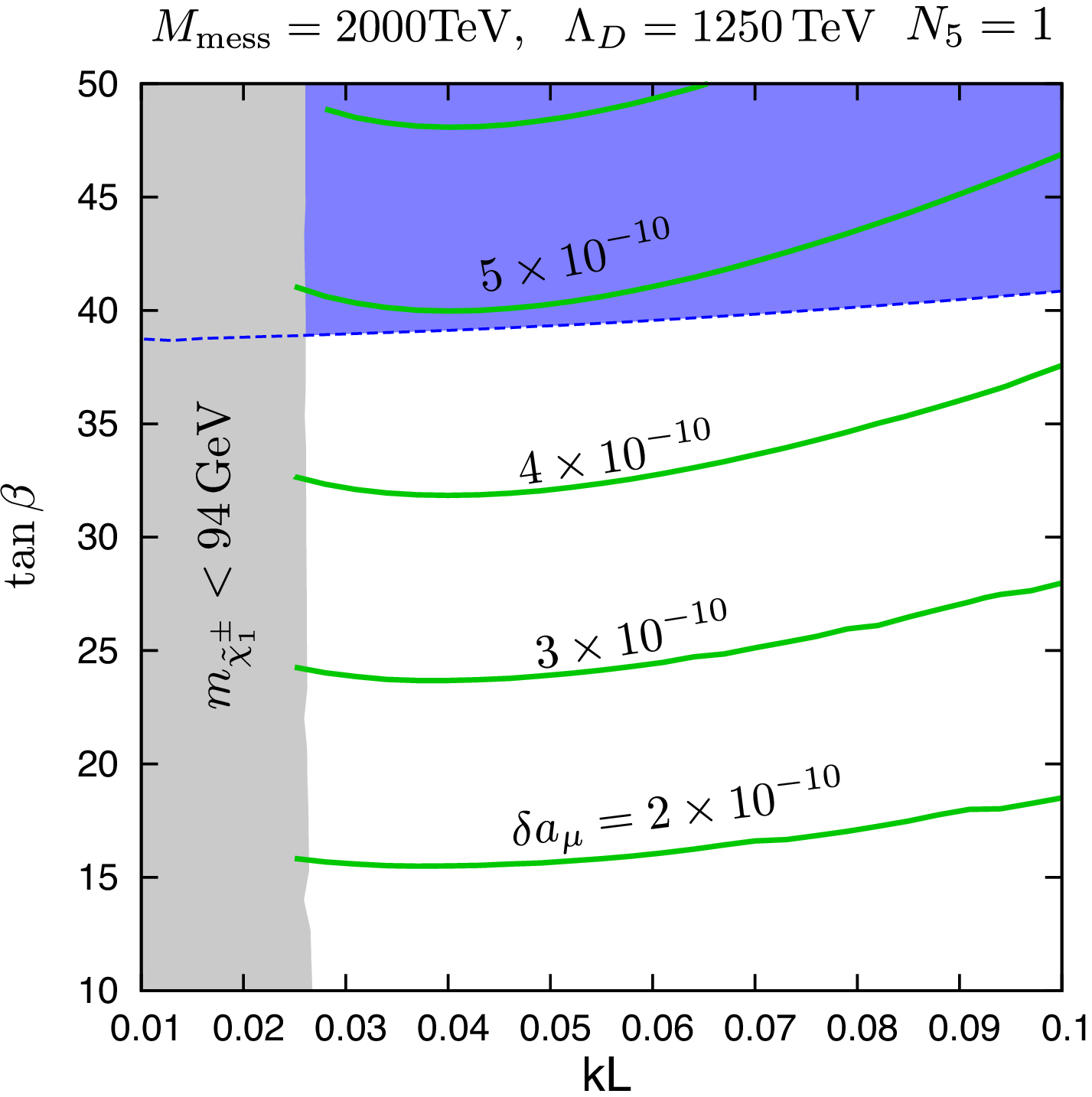}
\caption{\small \sl Contours of $\delta a_\mu$ on the $k_L$-$\tan\beta$ plane for $\Lambda_D = $ 1000 (left) and 1250 TeV (right). The region above the dashed blue line is excluded because of the unstable electroweak symmetry minimum. The region limited by the chargino mass bound is also shown~\cite{PDG}.}
\label{fig:5_1}
\end{center}
\end{figure}
%%%%%%%%%%%%%%%%%%%%%%%%%%%%%%%%%%%%%%%%%%%%%%%%%%%%%%%%%%%%

We next consider models with messengers in other than the adjoint representation. Before going to introduce the additional Higgs soft masses other than the gauge mediated SUSY breaking effects,
let us first see how it is difficult to explain the 125GeV Higgs mass and the deviation of
the muon anomalous magnetic simultaneously in those representations.
 In figure~\ref{fig:5_1}, the parameter space of the model with messengers in fundamental representations is shown. Here, we parameterize the splitting of the F-term by $k_L$,
\begin{eqnarray}
W
= (M_D +     F \theta^2) \Psi_D \overline{\Psi}_D
+ (M_L + k_L F \theta^2) \Psi_L \overline{\Psi}_L,
\end{eqnarray}
where $\overline{\Psi}_{D}$ and $\Psi_L$ are messenger multiplets which are transformed as $(\bar{\bf 3}, {\bf 1}, 1/3)$ and $({\bf 1}, {\bf 2}, -1/2)$ under the SM gauge groups, respectively. As in the case of the adjoint messenger, we take a common SUSY mass, i.e., $M_L = M_D = M_{\rm mess} = 2 \times 10^6$ GeV. In the left (right) panel of the figure, $\Lambda_D \equiv F/M_D$ is taken to be $\Lambda_D =$ 1000 (1250) TeV, which corresponds to the stop mass of $m_{\tilde{t}} \simeq$ 8 (10) TeV. The bino and higgsino masses turn out to be $M_1 \simeq$ 600 (800) GeV and $\mu \simeq$ 4000 (4900) GeV, respectively, in the left (right) panel. Since neither bino nor higgsino are light, $\delta a_{\rm \mu}$ cannot reach the value of $\sim 10^{-9}$; the observed value of the muon $g-2$ cannot be explained even with the splitting F-terms.
%The situation can be remedied if there are additional contributions to the Higgs mass squared, as shown below.

In the case of messengers in anti-symmetric representations of $SU(5)$, i.e., ${\bf 10} + \overline{\bf 10}$, the messenger multiplets $\Psi_Q$ ($SU(2)_L$ doublet) and $\Psi_U$ ($SU(2)_L$ singlet) have color charges. When the squark masses are generated dominantly from $\Psi_U$, the bino and slepton masses are heavier than those of the fundamental messenger case due to the larger hyper-charge of $\Psi_U$. When the  squark masses are dominantly generated by $\Psi_Q$, the left-handed sleptons and wino cannot be light, because $\Psi_Q$ has a $SU(2)_L$ charge. As a result, it is more difficult to enhance $\delta a_\mu$ in the ${\bf 10} + \overline{\bf 10}$ case.

Now, let us introduce additional soft-mass squared to the Higgs doublets in the model with the messengers in the fundamental representation.
As we have mentioned in Introduction, the additional Higgs soft masses allow
the small $\mu$-term, which enhances the wino-higgisno contibutions to the $g-2$.
Such additional contributions can be generated if the Higgs
doublets couple to the SUSY breaking sector.
%In the model with messengers in fundamental representation, the difficulties discussed above can be resolved if there are additional contributions to Higgs soft-mass squared. 
%{\bf (Models to have the additional contributions should be discussed here by referring appropriate papers.)}
%\vspace*{5cm}
For example, let us take a SUSY breaking O'Reifeartaigh model coupling to the Higgs doublets via,
\begin{eqnarray}
W = m^2 Z + \frac{\kappa}{2} Z X^2 + M_{XY}X Y + \lambda X H_u H_d\ .
\end{eqnarray}
Here,  $Z$ is a SUSY breaking field with a non-vanishing $F$-term, i.e. $\langle F_Z \rangle = m^2$,
and $X$, $Y$ are the singlet fields.
With the above interactions, the Higgs doublets receive non-vanishing and positive soft-mass squared 
at the one-loop level,
\begin{eqnarray}
\delta m_{H_{u,d}}^2 = \frac{\lambda^2}{32\pi^2} \frac{\tilde{F}^2}{M_{XY}^2} \left( 1 + {\cal O}(\tilde{F}^2/M_{XY}^4)\right)\ ,
\end{eqnarray}
where $\tilde{F} = \kappa m^2$.
In this way, we obtain desired additional soft-mass squared of the Higgs doublets  by arranging parameters appropriately.%
\footnote{
In this model, neither the $\mu$-term nor $B$-term are generated through the above interactions,
which are forbidden by a $Z_4$ symmetry under which $X$ and $Y$ change signs.
It is also possible to consider models where the $\mu$-term is also generated while keeping the $B$-term
suppressed via the couplings between the Higgs doublets and O'Raifeartaigh model\,\cite{sweetspot}.
In such models with $\mu$-term generation, it is predicted that $\mu$-term is rather suppressed
than $\delta m_{H_{u,d}}$.
}

%%%%%%%%%%%%%%%%%%%%%%%%%%%%%%%%%%%%%%%%%%%%%%%%%%%%%%%%%%%%
\begin{figure}[t]
\begin{center}
\includegraphics[scale=0.6]{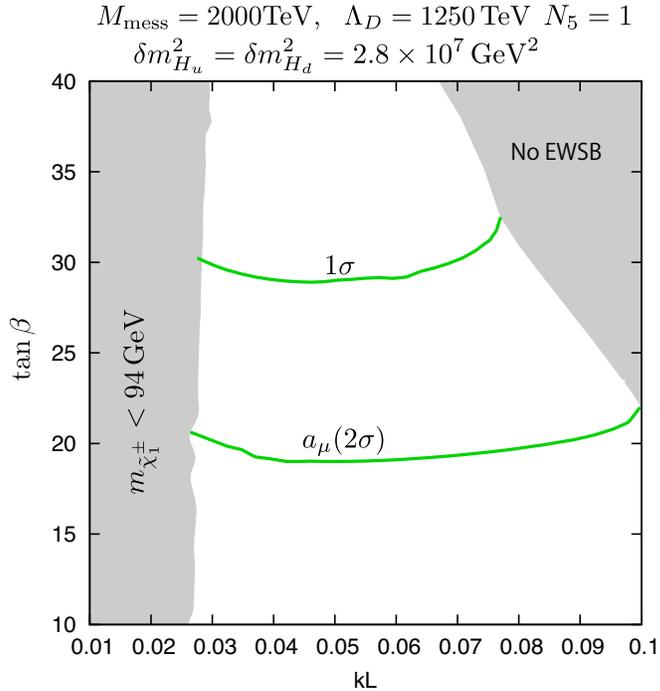}
\caption{\small \sl Contours of $a_\mu$ consistent with the experimental result at 1 and 2$\sigma$ C.L. on the $k_L$-$\tan\beta$ plane for $\Lambda_D = $ 1250 TeV. Additional contributions to $m_{H_u}^2$ and $m_{H_d}^2$ are taken to be $\delta m_{H_u}^2 = \delta m_{H_d}^2 =$ 2.8 $\times$ 10$^{\sl 7}$ GeV$^{\sl 2}$, respectively. The gray regions are excluded by chargino searches at the LEP experiment, or unsuccessful  electroweak symmetry breaking.}
\label{fig:5_2}
\end{center}
\end{figure}
%%%%%%%%%%%%%%%%%%%%%%%%%%%%%%%%%%%%%%%%%%%%%%%%%%%%%%%%%%%%

In the followings, we take the additional soft terms as free parameters and 
assume that they are generated at around the messenger scale. 
With the additional contributions to the Higgs soft square masses, $\delta m_{H_u}^2$ and $\delta m_{H_d}^2$, 
the $\mu$-parameter can be small (see Eq.(\ref{eq:mu})).
Thus, in this case, the higgsinos can be light as $\mathcal{O}(100)$ GeV, and hence,
%In following discussions, we simply assume the existence of the additional squared masses and the scale of their generations are treated as free parameters. As can be easily seen from Eq.~(\ref{eq:mu}), the $\mu$-parameter can be small if there is an large positive corrections to $m_{H_u}^2$. I
$\delta a_\mu$ can be large enough. In figure~{\ref{fig:5_2}}, contours of $a_\mu$ consistent with the experimental result at 1 and 2$\sigma$ C.L. are shown. The figure shows that the observed value of the anomalous magnetic moment can be explained within 1$\sigma$ level. The typical value of the $\mu$-parameter is less than about 300 GeV in the region consistent with the observed value within 2$\sigma$ level. In table \ref{table:55bar}, we show a mass spectrum and $\delta a_{\mu}$ as a reference. The bino and   sleptons are heavier than 700 GeV. On the other hand, the higgsino and wino are as light as 200 GeV and 160 GeV, respectively; $\delta a_{\mu}$ is enhanced with the help of the light higgsino and wino. The colored SUSY particles are as heavy as $8.6-11$ TeV, which are required to explain the Higgs boson mass of around  125 GeV.

In this model, the wino-like neutralino is the next-to-lightest SUSY particle. Since the gluino and squarks are as heavy as $\mathcal{O}(10)$ TeV, collider limits are much weaker than those of conventional SUSY models. At present, the most sensitive mode related to the wino search at the LHC experiment comes from the direct wino pair production associated with a jet~\cite{Direct wino production at LHC}. In this analysis, the disappearing charged track at inner detectors caused by the long-life of the charged wino (corresponding to the decay length of about 5\,cm when the wino-higgsino mixing is negligible) is utilized to reduce SM backgrounds. The transition radiation tracker (TRT) is currently used to find the track. Since the TRT is located about 1m away from the beam pipe, the limit on the wino mass is rather weak, i.e., the limit is $m_{\rm wino} >$ 100 GeV~\cite{ATLAS-CONF-2012-111}. The most inner detectors such as pixel and SCT detectors are planned to be used in near future, and then the wino mass up to about 500 GeV will be covered.

%%%%%%%%%%%%%%%%%%%%%%%%%%%%%%%%%%%%%%%%%%%%%
\begin{table}[t!]
  \begin{center}
    \begin{tabular}{  c | c  }
    $M_{\rm mess}$ & 2000 TeV \\
    $\Lambda_D$ & 1250 TeV \\
    $N_5$ & 1 \\
    $\tan\beta$ & 32 \\
    $k_L$ & 0.05 \\
    $\delta m_{H_{u,d}}^2$ & 2.8 $\times$ 10$^7$ GeV$^2$ \\ 
    \hline
\hline    
    $M_{\rm bino}$ & 827 GeV \\
    $M_{\rm wino}$ & 163 GeV \\
    $\mu$ & 211 GeV \\
\hline
    $m_{\rm gluino}$ & 8.6 TeV \\
    $m_{\tilde{t}}$ & 10.1 TeV \\
    $m_{\tilde{q}}$ & 11.0 TeV \\
    $m_{\tilde{e}_L} (m_{\tilde{\mu}_L})$ & 731 GeV\\
    $m_{\tilde{e}_R} (m_{\tilde{\mu}_R})$ & 1398 GeV \\
    $m_{\tilde{\tau}_1}$ & 592 GeV \\
    $m_{\chi_1^0}$ & 143 GeV \\
     $m_{\chi_1^{\pm}}$ & 145 GeV \\
    $ \delta a_{\mu}$ & 2.11 $\times$ 10$^{-9}$
    \end{tabular}
    \caption{\small \sl The mass spectrum and $\delta a_{\rm \mu}$. The bino mass, wino mass and higgsino mass parameters are also shown.
     }
  \label{table:55bar}
  \end{center}
\end{table}
%%%%%%%%%%%%%%%%%%%%%%%%%%%%%%%%%%%%%%%%%%%%%

%%%%%%%%%%%%%%%%%%%
%%%%% Summary %%%%%
%%%%%%%%%%%%%%%%%%%
\section{Summary}

In this paper, we have discussed whether or not the 125 GeV Higgs boson mass and the discrepancy of the magnetic moment $a_\mu$ can be simultaneously explained in models with heavy squarks and light sleptons. Such a simple possibility turns out to be highly constrained and we found that models with adjoint ${\bf 24}$ messengers are the unique possibility. In these models, the discrepancy of $a_\mu$ between theoretical prediction and experimental result can be reduced to around 1\,$\sigma$ C.L. for the stop mass of about $8-10$ TeV. In such a region consistent with the experimental result of the muon $g-2$, the stau is the next to the lightest SUSY particle with the mass of less than about 200 GeV. In the light of the  constraint from the LHC experiments, such a light stau should decay promptly via a R-parity violating operator (lepton number violating operator). It should be emphasized that the gravitino is still a viable candidate for dark matter because the violation of the R-parity is small.

On the other hand, once we admit additional contributions to Higgs soft masses, we found that models with   the messengers in the fundamental ${\bf 5}+\bar{\bf 5}$ representations can also explain the 125 GeV Higgs boson mass and the experimental result of the anomalous magnetic moment simultaneously. In these models, the predicted $a_\mu$ can be consistent with its experimental result within $1\sigma$ C.L. even for the stop mass of about 10 TeV. Although the additional Higgs soft masses are required, these masses may be related to the origin of $\mu/B\mu$ term, that is, the solution for the $\mu/B\mu$ problem in gauge mediation models.

%%%%%%%%%%%%%%%%%%%%%%%%%%%
%%%%% Acknowledgments %%%%%
%%%%%%%%%%%%%%%%%%%%%%%%%%%
\section*{Acknowledgements}

This work is supported by the Grant-in-Aid for Scientific research from the Ministry of Education, Science, Sports, and Culture (MEXT), Japan, No. 24740151 (M.I.), No. 22244021 (S.M. and T.T.Y.), No. 23740169 (S.M.), and also by the World Premier International Research Center Initiative (WPI Initiative), MEXT, Japan. The work of N.Y. is supported in part by JSPS Research Fellowships for Young Scientists.

\appendix

%%%%%%%%%%%%%%%%%%%%%%
%%%%% Appendix A %%%%%
%%%%%%%%%%%%%%%%%%%%%%
\section{Gaugino and scalar masses}

Here, we summarize gaugino and scalar masses predicted by gauge mediation models with messenger multiplets in various representations. Note that $g_1$ is the coupling constant of $U(1)_Y$ gauge interaction with the GUT normalization

\subsection{${\bf 5}$+$\bar{\bf 5}$ messenger}

Gaugino masses are given by
\begin{eqnarray}
M_1 \simeq \frac{g_1^2}{16\pi^2} N_5
\left(\frac{2}{5} \Lambda_D + \frac{3}{5} \Lambda_L\right), \ \ 
M_2 \simeq \frac{g_2^2}{16\pi^2} N_5 (\Lambda_L), \ \ 
M_3 \simeq \frac{g_3^2}{16\pi^2} N_5 (\Lambda_D),
\end{eqnarray}
where $\Lambda_D = F_D/M_D$ and $\Lambda_L = F_L/M_L$. Here, sub-leading contributions are neglected. Scalar masses are, on the other hand, given by
\begin{eqnarray}
m_{\tilde{Q}}^2 &\simeq& N_5 \frac{2}{(16\pi^2)^2}
\left[
\frac43 g_3^4 \Lambda_D^2 + \frac34 g_2^4 \Lambda_L^2
+ \frac35 g_1^4 \left( \frac25 \Lambda_D^2 + \frac35 \Lambda_L^2 \right)
\frac{1}{6^2}
\right],
\nonumber \\
m_{\tilde{U}}^2 &\simeq& N_5 \frac{2}{(16\pi^2)^2}
\left[
\frac43 g_3^4 \Lambda_D^2
+ \frac35 g_1^4 \left( \frac25 \Lambda_D^2 + \frac35 \Lambda_L^2 \right)
\left(\frac23\right)^2
\right],
\nonumber \\
m_{\tilde{D}}^2 &\simeq& N_5 \frac{2}{(16\pi^2)^2}
\left[\frac43 g_3^4 \Lambda_D^2
+ \frac35 g_1^4 \left( \frac25 \Lambda_D^2 + \frac35 \Lambda_L^2 \right)
\frac{1}{3^2}
\right],
\nonumber \\
m_{\tilde{L}}^2 &\simeq& N_5 \frac{2}{(16\pi^2)^2}
\left[
\frac34 g_2^4 \Lambda_L^2
+ \frac35 g_1^4 \left( \frac25 \Lambda_D^2 + \frac35 \Lambda_L^2 \right)
\frac{1}{2^2}
\right],
\nonumber \\
m_{\tilde{E}}^2 &\simeq& N_5 \frac{2}{(16\pi^2)^2}
\left[
\frac35 g_1^4 \left( \frac25 \Lambda_D^2 + \frac35 \Lambda_L^2 \right)
\right],
\nonumber \\
m_{H_u}^2 &=& m_{H_d}^2 = m_{\tilde{L}}^2.
\end{eqnarray}

\subsection{${\bf 10}$ + $\overline{{\bf 10}}$ messenger}

Gaugino masses are given by
\begin{eqnarray}
M_1 &\simeq& \frac{g_1^2}{16\pi^2}
\left(\frac15 \Lambda_Q + \frac85 \Lambda_U + \frac65 \Lambda_E\right), \ \ 
M_2 \simeq \frac{g_2^2}{16\pi^2} (3 \Lambda_Q), \nonumber \\ 
M_3 &\simeq& \frac{g_3^2}{16\pi^2} (2 \Lambda_Q + \Lambda_U),
\end{eqnarray}
where $\Lambda_Q = F_Q/M_Q$, $\Lambda_U = F_U/M_U$ and $\Lambda_E = F_E/M_E$. Scalar masses are given by
\begin{eqnarray}
m_{\tilde{Q}}^2 &\simeq& \frac{2}{(16\pi^2)^2}
\left[\frac43 g_3^4 \tilde{\Lambda}_3^2 + \frac34 g_2^4 (3 \Lambda_Q^2)
+ \frac35 g_1^4 (\tilde{\Lambda}_1^2) \frac{1}{6^2}
\right],
\nonumber \\
m_{\tilde{U}}^2 &\simeq& \frac{2}{(16\pi^2)^2}
\left[
\frac43 g_3^4 \tilde{\Lambda}_3^2 + \frac35 g_1^4 \tilde{\Lambda}_1^2
\left(\frac23\right)^2
\right],
\nonumber \\
m_{\tilde{D}}^2 &\simeq& \frac{2}{(16\pi^2)^2}
\left[\frac43 g_3^4 \tilde{\Lambda}_3^2 + \frac35 g_1^4 \tilde{\Lambda}_1^2
\frac{1}{3^2}
\right],
\nonumber \\
m_{\tilde{L}}^2 &\simeq& \frac{2}{(16\pi^2)^2}
\left[ \frac34 g_2^4 (3 \Lambda_Q^2) + \frac35 g_1^4 \tilde{\Lambda}_1^2
\frac{1}{2^2}
\right],
\nonumber \\
m_{\tilde{E}}^2 &\simeq& \frac{2}{(16\pi^2)^2}
\left[ \frac35 g_1^4 \tilde{\Lambda}_1^2 \right],
\nonumber \\
m_{H_u}^2 &=& m_{H_d}^2 = m_{\tilde{L}}^2,
\end{eqnarray}
where $\tilde{\Lambda}_1^2 = (\Lambda_Q^2 + 8 \Lambda_U^2 + 6 \Lambda_E^2)/5$ and $\tilde{\Lambda}_3^2= 2 \Lambda_Q^2 + \Lambda_U^2$, respectively.

\subsection{Adjoint messenger}
Gaugino masses are given by
\begin{eqnarray}
M_1 \simeq \frac{g_1^2}{16\pi^2} (5 \Lambda_X), \ \ 
M_2 \simeq \frac{g_2^2}{16\pi^2} (2 \Lambda_3 + 3 \Lambda_X), \ \ 
M_3 \simeq \frac{g_3^2}{16\pi^2} (3 \Lambda_8 + 2 \Lambda_X),
\end{eqnarray}
where $\Lambda_8 = F_8/M_8$, $\Lambda_3 = F_3/M_3$ and $\Lambda_X=F_X/M_X$. Scalar masses are given by
\begin{eqnarray}
m_{\tilde{Q}}^2 &\simeq& \frac{2}{(16\pi^2)^2}
\left[
\frac43 g_3^4 (3\Lambda_8^2 + 2 \Lambda_X^2)
+ \frac34 g_2^4 (2\Lambda_3^2 + 3 \Lambda_X^2)
+ \frac35 g_1^4 (5 \Lambda_X^2) \frac{1}{6^2}
\right],
\nonumber \\
m_{\tilde{U}}^2 &\simeq& \frac{2}{(16\pi^2)^2}
\left[
\frac43 g_3^4 (3\Lambda_8^2 + 2 \Lambda_X^2) + \frac35 g_1^4 (5 \Lambda_X^2)
\left(\frac23\right)^2
\right],
\nonumber \\
m_{\tilde{D}}^2 &\simeq& \frac{2}{(16\pi^2)^2}
\left[\frac43 g_3^4 (3\Lambda_8^2 + 2 \Lambda_X^2)  + \frac35 g_1^4 (5 \Lambda_X^2) \frac{1}{3^2}
\right],
\nonumber \\
m_{\tilde{L}}^2 &\simeq& \frac{2}{(16\pi^2)^2}
\left[ \frac34 g_2^4 (2\Lambda_3^2 + 3 \Lambda_X^2) + \frac35 g_1^4 (5 \Lambda_X^2)
\frac{1}{2^2}
\right],
\nonumber \\
m_{\tilde{E}}^2 &\simeq& \frac{2}{(16\pi^2)^2}
\left[ \frac35 g_1^4 (5 \Lambda_X^2) \right],
\nonumber \\
m_{H_u}^2 &=& m_{H_d}^2 = m_{\tilde{L}}^2.
\end{eqnarray}

%\vspace{-21pt}
%%%%%%%%%%%%%%%%%%%%%%%%
%%%%% Bibliography %%%%%
%%%%%%%%%%%%%%%%%%%%%%%%


\begin{thebibliography}{99}

%%%%%%%%%%%
%% Intro %%
%%%%%%%%%%%
\bibitem{ATLAS_Higgs}
G.~Aad {\it et al.}  [ATLAS Collaboration],
%``Observation of a new particle in the search for the Standard Model Higgs boson with the ATLAS detector at the LHC,''
Phys.\ Lett.\ B {\bf 716}, 1 (2012).
%[arXiv:1207.7214 [hep-ex]].

\bibitem{CMS_Higgs}
S.~Chatrchyan {\it et al.}  [CMS Collaboration],
%``Observation of a new boson at a mass of 125 GeV with the CMS experiment at the LHC,''
Phys.\ Lett.\ B {\bf 716}, 30 (2012).
%[arXiv:1207.7235 [hep-ex]].

%\cite{Okada:1990gg}
\bibitem{OYY1} 
%\cite{Okada:1990vk}
  Y.~Okada, M.~Yamaguchi and T.~Yanagida,
  %``Upper bound of the lightest Higgs boson mass in the minimal supersymmetric standard model,''
  Prog.\ Theor.\ Phys.\  {\bf 85}, 1 (1991);
    J.~R.~Ellis, G.~Ridolfi and F.~Zwirner,
  %``Radiative corrections to the masses of supersymmetric Higgs bosons,''
  Phys.\ Lett.\ B {\bf 257}, 83 (1991);
  %%CITATION = PHLTA,B257,83;%%
%\cite{Ellis:1991zd}
%\bibitem{Ellis:1991zd} 
%  J.~R.~Ellis, G.~Ridolfi and F.~Zwirner,
  %``On radiative corrections to supersymmetric Higgs boson masses and their implications for LEP searches,''
%  Phys.\ Lett.\ B {\bf 262}, 477 (1991);
  %%CITATION = PHLTA,B262,477;%%
%\cite{Haber:1990aw}
%\bibitem{Haber:1990aw} 
  H.~E.~Haber and R.~Hempfling,
  %``Can the mass of the lightest Higgs boson of the minimal supersymmetric model be larger than m(Z)?,''
  Phys.\ Rev.\ Lett.\  {\bf 66}, 1815 (1991).
  %%CITATION = PRLTA,66,1815;%%

  
 \bibitem{OYY2}
  %%CITATION = PTPKA,85,1;%%
  Y.~Okada, M.~Yamaguchi and T.~Yanagida,
  %``Renormalization group analysis on the Higgs mass in the softly broken supersymmetric standard model,''
  Phys.\ Lett.\ B {\bf 262}, 54 (1991).
  %%CITATION = PHLTA,B262,54;%%
  


  \bibitem{PureGM} 
  M.~Ibe and T.~T.~Yanagida,
  %``The Lightest Higgs Boson Mass in Pure Gravity Mediation Model,''
  Phys.\ Lett.\ B {\bf 709}, 374 (2012)
  [arXiv:1112.2462 [hep-ph]];
  %%CITATION = ARXIV:1112.2462;%%
  M.~Ibe, S.~Matsumoto and T.~T.~Yanagida,
  %``Pure Gravity Mediation with m_{3/2} = 10-100TeV,''
  Phys.\ Rev.\ D {\bf 85}, 095011 (2012)
  [arXiv:1202.2253 [hep-ph]];
  %%CITATION = ARXIV:1202.2253;%%
  %\cite{Bhattacherjee:2012ed}
%\bibitem{Bhattacherjee:2012ed} 
  B.~Bhattacherjee, B.~Feldstein, M.~Ibe, S.~Matsumoto and T.~T.~Yanagida,
  %``Pure Gravity Mediation of Supersymmetry Breaking at the LHC,''
  arXiv:1207.5453 [hep-ph].
  %%CITATION = ARXIV:1207.5453;%%


\bibitem{extramatter} 
  T.~Moroi and Y.~Okada,
  %``Radiative corrections to Higgs masses in the supersymmetric model with an extra family and antifamily,''
  Mod.\ Phys.\ Lett.\ A {\bf 7}, 187 (1992);
T.~Moroi and Y.~Okada,
  %``Upper bound of the lightest neutral Higgs mass in extended supersymmetric Standard Models,''
  Phys.\ Lett.\ B {\bf 295}, 73 (1992);
  K.~S.~Babu, I.~Gogoladze, M.~U.~Rehman and Q.~Shafi,
  %``Higgs Boson Mass, Sparticle Spectrum and Little Hierarchy Problem in Extended MSSM,''
  Phys.\ Rev.\ D {\bf 78}, 055017 (2008)
  [arXiv:0807.3055 [hep-ph]];
  %%CITATION = ARXIV:0807.3055;%%;
  S.~P.~Martin,
  %``Extra vector-like matter and the lightest Higgs scalar boson mass in low-energy supersymmetry,''
  Phys.\ Rev.\ D {\bf 81}, 035004 (2010)
  [arXiv:0910.2732 [hep-ph]].
  %%CITATION = ARXIV:0910.2732;%%
%  M.~Asano, T.~Moroi, R.~Sato and T.~T.~Yanagida,
  %``Non-anomalous Discrete R-symmetry, Extra Matters, and Enhancement of the Lightest SUSY Higgs Mass,''
%  Phys.\ Lett.\ B {\bf 705}, 337 (2011)
%  [arXiv:1108.2402 [hep-ph]].
  %%CITATION = ARXIV:1108.2402;%%


\bibitem{extragauge}
% extra gauge interaction
%\bibitem{Batra:2003nj} 
  P.~Batra, A.~Delgado, D.~E.~Kaplan and T.~M.~P.~Tait,
  %``The Higgs mass bound in gauge extensions of the minimal supersymmetric standard model,''
  JHEP {\bf 0402}, 043 (2004)
  [hep-ph/0309149];
  %%CITATION = HEP-PH/0309149;%%
%\cite{Maloney:2004rc}
%\bibitem{Maloney:2004rc} 
  A.~Maloney, A.~Pierce and J.~G.~Wacker,
  %``D-terms, unification, and the Higgs mass,''
  JHEP {\bf 0606}, 034 (2006)
  [hep-ph/0409127].
  %%CITATION = HEP-PH/0409127;%%


\bibitem{NMSSM1}
  M.~Maniatis,
  %``The Next-to-Minimal Supersymmetric extension of the Standard Model reviewed,''
  Int.\ J.\ Mod.\ Phys.\ A {\bf 25}, 3505 (2010)
  [arXiv:0906.0777 [hep-ph]].
  %%CITATION = ARXIV:0906.0777;%%
%\cite{Ellwanger:2009dp}
\bibitem{NMSSM2} 
  U.~Ellwanger, C.~Hugonie and A.~M.~Teixeira,
  %``The Next-to-Minimal Supersymmetric Standard Model,''
  Phys.\ Rept.\  {\bf 496}, 1 (2010)
  [arXiv:0910.1785 [hep-ph]].
  %%CITATION = ARXIV:0910.1785;%%

%\bibitem{Drastic}
%\cite{Evans:2012uf}
%\bibitem{Evans:2012uf} 
%  J.~L.~Evans, M.~Ibe and T.~T.~Yanagida,
  %``The Lightest Higgs Boson Mass in the MSSM with Strongly Interacting Spectators,''
%  Phys.\ Rev.\ D {\bf 86}, 015017 (2012)
%  [arXiv:1204.6085 [hep-ph]].
  %%CITATION = ARXIV:1204.6085;%%

\bibitem{EIY_gm2}
J.~L.~Evans, M.~Ibe, S.~Shirai and T.~T.~Yanagida,
%``A 125GeV Higgs Boson and Muon g-2 in More Generic Gauge Mediation,''
Phys.\ Rev.\ D {\bf 85}, 095004 (2012).
%[arXiv:1201.2611 [hep-ph]].

\bibitem{EXT_gm2}
M.~Endo, K.~Hamaguchi, S.~Iwamoto and N.~Yokozaki,
%``Higgs Mass and Muon Anomalous Magnetic Moment in Supersymmetric Models with Vector-Like Matters,''
Phys.\ Rev.\ D {\bf 84}, 075017 (2011);
%[arXiv:1108.3071 [hep-ph]].
M.~Endo, K.~Hamaguchi, S.~Iwamoto and N.~Yokozaki,
%``Higgs mass, muon g-2, and LHC prospects in gauge mediation models with vector-like matters,''
Phys.\ Rev.\ D {\bf 85}, 095012 (2012).
%[arXiv:1112.5653 [hep-ph]].

%\cite{Nakayama:2012zc}
\bibitem{PQ_EXT} 
  K.~Nakayama, N.~Yokozaki and N.~Yokozaki,
  %``Peccei-Quinn extended gauge-mediation model with vector-like matter,''
  arXiv:1204.5420 [hep-ph].
  %%CITATION = ARXIV:1204.5420;%%

\bibitem{EXT_gauge_gm2}
  M.~Endo, K.~Hamaguchi, S.~Iwamoto, K.~Nakayama and N.~Yokozaki,
  %``Higgs mass and muon anomalous magnetic moment in the U(1) extended MSSM,''
  Phys.\ Rev.\ D {\bf 85}, 095006 (2012)
  [arXiv:1112.6412 [hep-ph]].
  %%CITATION = ARXIV:1112.6412;%%

%%%%%%%%%
%% g-2 %%
%%%%%%%%%
\bibitem{Bennett:2006fi}
G.~W.~Bennett {\it et al.}  [Muon G-2 Collaboration],
%``Final Report of the Muon E821 Anomalous Magnetic Moment Measurement at BNL,''
Phys.\ Rev.\ D {\bf 73}, 072003 (2006).
%[hep-ex/0602035].

\bibitem{Hagiwara:2011af}
K.~Hagiwara, R.~Liao, A.~D.~Martin, D.~Nomura and T.~Teubner,
%``(g-2)_mu and alpha(M_Z^2) re-evaluated using new precise data,''
J.\ Phys.\ G G {\bf 38}, 085003 (2011).
%[arXiv:1105.3149 [hep-ph]].

\bibitem{EIY1}
J.~L.~Evans, M.~Ibe and T.~T.~Yanagida,
%``Relatively Heavy Higgs Boson in More Generic Gauge Mediation,''
Phys.\ Lett.\ B {\bf 705}, 342 (2011).
%[arXiv:1107.3006 [hep-ph]].

\bibitem{gm2_MSSM}
G.~-C.~Cho, K.~Hagiwara, Y.~Matsumoto and D.~Nomura,
%``The MSSM confronts the precision electroweak data and the muon g-2,''
JHEP {\bf 1111}, 068 (2011).
%[arXiv:1104.1769 [hep-ph]].

%%%%%%%%%%%%%
%% Adjoint %%
%%%%%%%%%%%%%
\bibitem{suspect}
A.~Djouadi, J.~-L.~Kneur and G.~Moultaka,
%``SuSpect: A Fortran code for the supersymmetric and Higgs particle spectrum in the MSSM,''
Comput.\ Phys.\ Commun.\  {\bf 176}, 426 (2007).
%[hep-ph/0211331].

\bibitem{FeynHiggs}
S.~Heinemeyer, W.~Hollik and G.~Weiglein,
%``FeynHiggs: A Program for the calculation of the masses of the neutral CP even Higgs bosons in the MSSM,''
Comput.\ Phys.\ Commun.\ \ {\bf 124}, 76  (2000);
%[hep-ph/9812320].
%\bibitem{hep-ph/9812472}
S.~Heinemeyer, W.~Hollik and G.~Weiglein,
%``The Masses of the neutral CP - even Higgs bosons in the MSSM: Accurate analysis at the two loop level,''
Eur.\ Phys.\ J.\ C\ {\bf 9}, 343  (1999);
%[hep-ph/9812472].
%\bibitem{hep-ph/0212020}
G.~Degrassi, S.~Heinemeyer, W.~Hollik, P.~Slavich and G.~Weiglein,
%``Towards high precision predictions for the MSSM Higgs sector,''
Eur.\ Phys.\ J.\ C\ {\bf 28}, 133  (2003);
%[hep-ph/0212020].
%\bibitem{hep-ph/0611326}
%M.~Frank, T.~Hahn, S.~Heinemeyer, W.~Hollik, H.~Rzehak and G.~Weiglein,
%``The Higgs Boson Masses and Mixings of the Complex MSSM in the Feynman-Diagrammatic Approach,''
JHEP\ {\bf 0702}, 047  (2007).
%[hep-ph/0611326].


\bibitem{Rattazzi:1996fb}
R.~Rattazzi and U.~Sarid,
%``Large tan Beta in gauge mediated SUSY breaking models,''
Nucl.\ Phys.\ B {\bf 501}, 297 (1997).
%[hep-ph/9612464].

\bibitem{Hisano:2010re}
J.~Hisano and S.~Sugiyama,
%``Charge-breaking constraints on left-right mixing of stau's,''
Phys.\ Lett.\ B {\bf 696}, 92 (2011).
%[arXiv:1011.0260 [hep-ph]].


\bibitem{PDG}
J. Beringer et al. (Particle Data Group), Phys. Rev. D86, 010001 (2012)

\bibitem{ATLAS_stau1}
The ATLAS Collaboration,
ATLAS-CONF-2012-075.

\bibitem{LFV_models}
L.~J.~Hall and M.~Suzuki,
%``Explicit R-Parity Breaking in Supersymmetric Models,''
Nucl.\ Phys.\ B {\bf 231}, 419 (1984);
%\bibitem{Brahm:1989iy}
D.~E.~Brahm and L.~J.~Hall,
%``Low-energy Lepton Violation From Supersymmetric Flipped Su(5),''
Phys.\ Rev.\ D {\bf 40}, 2449 (1989);
W.~Buchmuller, L.~Covi, K.~Hamaguchi, A.~Ibarra and T.~Yanagida,
%``Gravitino Dark Matter in R-Parity Breaking Vacua,''
JHEP {\bf 0703}, 037 (2007).
%[hep-ph/0702184 [HEP-PH]].

\bibitem{RPVLEP}
%\bibitem{Heister:2002jc}
A.~Heister {\it et al.}  [ALEPH Collaboration],
%``Search for supersymmetric particles with R parity violating decays in $e^{+} e^{-}$ collisions at $\sqrt{s}$ up to 209-GeV,''
Eur.\ Phys.\ J.\ C {\bf 31}, 1 (2003);
%[hep-ex/0210014].
%\bibitem{Braibant:2003px} 
S.~Braibant,
%``SUSY searches at LEP,''
hep-ex/0305058.

\bibitem{washout}
%\bibitem{RpVcosmobound}
%\cite{Campbell:1990fa}
%\bibitem{Campbell:1990fa}
  B.~A.~Campbell, S.~Davidson, J.~R.~Ellis and K.~A.~Olive,
  %``Cosmological baryon asymmetry constraints on extensions of the standard
  %model,''
  Phys.\ Lett.\  B {\bf 256} (1991) 457; %%%%% Note!.... SPIRES record is wrong .....
%  Phys.\ Lett.\  B {\bf 256} (1991) 484.
  %%CITATION = PHLTA,B256,484;%%
%%%%%  
%\cite{Fischler:1990gn}
%\bibitem{Fischler:1990gn}
  W.~Fischler, G.~F.~Giudice, R.~G.~Leigh and S.~Paban,
  %``Constraints On The Baryogenesis Scale From Neutrino Masses,''
  Phys.\ Lett.\  B {\bf 258} (1991) 45; 
  %%CITATION = PHLTA,B258,45;%%
%%%%%
%\cite{Dreiner:1992vm}
%\bibitem{Dreiner:1992vm}
  H.~K.~Dreiner and G.~G.~Ross,
  %``Sphaleron Erasure Of Primordial Baryogenesis,''
  Nucl.\ Phys.\  B {\bf 410} (1993) 188
  [arXiv:hep-ph/9207221].
  %%CITATION = NUPHA,B410,188;%%

\bibitem{LFV_LNV_ENDO}
M.~Endo, K.~Hamaguchi and S.~Iwamoto,
%``Lepton Flavor Violation and Cosmological Constraints on R-parity Violation,''
JCAP {\bf 1002}, 032 (2010).
%[arXiv:0912.0585 [hep-ph]].

\bibitem{Matsumoto:2011fk}
S.~Matsumoto and T.~Moroi,
%``Studying Very Light Gravitino at the ILC,''
Phys.\ Lett.\ B {\bf 701}, 422 (2011).
%[arXiv:1104.3624 [hep-ph]].


\bibitem{sweetspot}
  M.~Ibe and R.~Kitano,
  %``Sweet Spot Supersymmetry,''
  JHEP {\bf 0708}, 016 (2007)
  [arXiv:0705.3686 [hep-ph]].
  %%CITATION = ARXIV:0705.3686;%%

%%%%%%%%%%%%
%% Others %%
%%%%%%%%%%%%
\bibitem{Direct wino production at LHC}
%\bibitem{Feng:1999fu}
J.~L.~Feng, T.~Moroi, L.~Randall, M.~Strassler and S.~-f.~Su,
%``Discovering supersymmetry at the Tevatron in wino LSP scenarios,''
Phys.\ Rev.\ Lett.\  {\bf 83}, 1731 (1999);
%[hep-ph/9904250].
%\bibitem{Ibe:2006de}
M.~Ibe, T.~Moroi and T.~T.~Yanagida,
%``Possible Signals of Wino LSP at the Large Hadron Collider,''
Phys.\ Lett.\ B {\bf 644}, 355 (2007);
%[hep-ph/0610277].
%\bibitem{Moroi:2011ab}
T.~Moroi and K.~Nakayama,
%``Wino LSP detection in the light of recent Higgs searches at the LHC,''
Phys.\ Lett.\ B {\bf 710}, 159 (2012);
%[arXiv:1112.3123 [hep-ph]].
%\bibitem{Bhattacherjee:2012ed}
B.~Bhattacherjee, B.~Feldstein, M.~Ibe, S.~Matsumoto and T.~T.~Yanagida,
%``Pure Gravity Mediation of Supersymmetry Breaking at the LHC,''
arXiv:1207.5453 [hep-ph].

\bibitem{ATLAS-CONF-2012-111} 
The ATLAS Collaboration,
%``Search for direct chargino production in anomaly-mediated supersymmetry breaking models based on a disappearing-track signature, in pp collisions at $\sqrt{s} =$ 7 TeV with the ATLAS detector at the LHC,''
ATLAS-CONF-2012-111.

\end{thebibliography}
\end{document}